# Visualization of spin-orbit entangled 4f electrons in crystalline materials

Shunsuke Kitou[1]*, Kentaro Ueda[2], Yuiga Nakamura[3], Kunihisa Sugimoto[4], Yusuke Nomura[5,6], Ryotaro Arita[7,8], Yoshinori Tokura[2,8,9], Taka-hisa Arima[1,8]

[1]Department of Advanced Materials Science, The University of Tokyo, Kashiwa 277-8561, Japan

[2]Department of Applied Physics, The University of Tokyo, Tokyo 113-8656, Japan

[3]Japan Synchrotron Radiation Research Institute (JASRI), SPring-8, Hyogo 679-5198, Japan

[4]Department of Chemistry, Kindai University, Osaka 577-8502, Japan

[5]Institute for Materials Research, Tohoku University, Sendai, 980-8577, Japan

[6]Advanced Institute for Materials Research, Tohoku University, Sendai, 980-8577 Japan

[7]Department of Physics, The University of Tokyo, Tokyo 113-0033, Japan

[8]RIKEN Center for Emergent Matter Science (CEMS), Wako 351-0198, Japan

[9]Tokyo College, The University of Tokyo, Tokyo 113-8656, Japan

*Shunsuke Kitou

**Email:** kitou@edu.k.u-tokyo.ac.jp

**Author Contributions:** S.K. and T.-h.A. designed research; K.U. and Y.T. grew the crystal; S.K., Y. Nakamura, and K.S. performed the XRD experiment; S.K. analyzed the XRD data; Y. Nomura and R.A. performed the DFT calculation; and S.K. and T.-h.A. wrote the paper.

**Competing Interest Statement:** The authors declare no competing interests.

## Abstract

Lanthanide 4f electrons are strongly influenced by spin–orbit coupling, resulting in well-defined *J* multiplets, which are further split by the crystalline electric field in condensed matter. While the anisotropy of 4f electrons is closely linked to material properties, direct experimental observation of the 4f electron distribution in real space remains a significant challenge. Here, we present an approach for visualizing the anisotropic distribution of lanthanide 4f electrons in pyrochlore oxides by combining high-photon-energy X-ray diffraction and valence electron density (VED) analysis based on the core differential Fourier synthesis (CDFS) method. The observed VED distributions around the lanthanide site reveal the parameters of the ground-state wavefunction, which roughly agree with point-charge calculations for the trigonal crystal electric field under the *LS* coupling scheme. This CDFS-based VED observation method not only provides insights into the anisotropic nature of 4f electrons but also opens a pathway for studying the 4f states in a wide range of crystalline materials.

## Main Text

## Introduction

Lanthanide compounds not only demonstrate high performance as applied materials—such as permanent magnets [1] and phosphors [2]—but also have the potential to become next-generation innovative materials in the fields of spintronics [3] and quantum computing [4,5]. Moreover, lanthanide compounds offer a playground of interesting physical properties, including diverse magnetic structures [6], multipole order [7], and unconventional superconductivity [8,9]. Since 5s and 5p electrons partly shield the effect of surrounding atoms, 4f electrons on the lanthanide atom hardly participate in chemical bonding with surrounding ions and form $J$ multiplets owing to the strong relativistic spin-orbit coupling (SOC). Crystalline electric field (CEF) from surrounding ions further splits the multiplets. The resultant eigenstates have anisotropic spatial distribution and hence can host various multipoles. The wavefunctions and energy levels of the 4f states, determined by SOC and CEF, not only govern the material's magnetization and specific heat but also influence various properties such as magnetic anisotropy, Kondo effect [10], and multipole order/fluctuation [7].

If the spatial distribution of 4f electrons can be visualized, it enables direct understanding of the 4f-electron wavefunction. The 3d electrons in transition metal compounds have been observed by using several techniques [11-15]. In contrast, the methodologies for the experimental observation of 4f electrons are limited [16]. Although inelastic neutron scattering has been the most commonly used method to analyze the 4f level scheme, the wavefunctions estimated by fitting the scattering intensity involve significant ambiguity. X-ray diffraction is a good candidate for the observation of the spatial distribution of electrons. In fact, the spatial distribution of 3d electrons was successfully observed by cutting-edge analysis of X-ray scattering. For example, the spatial distribution of valence electron density (VED) in various crystalline materials has been visualized by combining high-photon-energy X-ray diffraction and a core differential Fourier synthesis (CDFS) analysis [15,17-23]. Nonetheless, the direct observation of the spatial distribution of 4f electrons remains a significant challenge because (i) 4f electrons exhibit steeper spatial modulation compared to 3d electrons, demanding higher spatial resolution, and (ii) lanthanide elements accommodate a larger number of core electrons than 3d transition elements, necessitating a wider dynamic range.

In this study, we visualize the 4f VED distribution on the lanthanide ions at the $A$ site in pyrochlore oxides $A_2Ir_2O_7$ ($A$ = Pr, Nd, and Eu) through CDFS analysis using single-crystal high-photon-energy X-ray diffraction. We demonstrate that the 4f state can be determined directly from the anisotropy of the VED distribution. High-photon-energy X-ray diffraction experiments at the synchrotron radiation facility SPring-8 guarantee a spatial resolution of approximately 0.25 Å and a dynamic range of intensity exceeding $10^6$. Furthermore, we provide experimental guidelines for observing 5d electrons using next-generation synchrotron radiation facilities.

**Results**

The pyrochlore iridates $A_2Ir_2O_7$ ($A$ = lanthanide) with cubic space group $Fd\bar{3}m$ [24,25] undergo a transition from an antiferromagnetic insulator/semimetal ($A$= Y, Dy, Gd, Eu, Sm, Nd) to a paramagnetic metal ($A$ = Pr) by replacing $A$ ions [26-29], where exotic electronic and magnetic properties such as anomalous Hall effect [30-32], magnetic-field-induced metal-insulator transition [33], spin-liquid behavior [34], and spin-ice-like magnetic order [35] are realized. Figure 1A shows the crystal structure of $A_2Ir_2O_7$. The $A$-ion dependence of the crystal structure of $A_2Ir_2O_7$ ($A$ = Eu, Nd, and Pr) at 100 K is summarized in Fig. S1, where the Ir–O–Ir bond angle governing the magnetic interactions between Ir moments is observed to change systematically with the $A$-ion. The $A$ and Ir atoms each form a pyrochlore network. While each Ir atom is surrounded by six O(1), which form a trigonally distorted octahedron, each $A$ atom is surrounded by eight O, six O(1) and two O(2), as shown in Fig. 1B. The local symmetry at both $A$ and Ir sites is $.\bar{3}m$ ($D_{3d}$). The formal valences of $A$ and Ir are +3 and +4, respectively. For example, in the case of $A^{3+}$ = $Pr^{3+}$ with $4f^2$ electrons, the orbital and spin angular momentum quantum numbers in the low-lying multiplets are $L$ = 5 and $S$ = 1, respectively. The total angular momentum quantum number in the ground state is $J = |L - S| = 4$ when considering the SOC under the $LS$ coupling scheme (Fig. 1C). In the trigonal CEF, the $J$ = 4 nonuplet $^3H_4$ is expected to split into three doublets and three singlets. Similarly, for $A^{3+}$ = $Nd^{3+}$ with $4f^3$ electrons, the $J$ = 9/2 decuplet $^4I_{9/2}$ should split into five Kramers doublets (Fig. 1D). Figures 1E and 1F show the $4f^2$ and $4f^3$ electron density distributions of eigenstates corresponding to $J_z$ = ±4, ±3, ±2, ±1, 0, and $J_z$ = ±9/2, ±7/2, ±5/2, ±3/2, ±1/2, respectively. Although the ground state in the CEF can be represented by a linear combination of the $J_z$ eigenstates, experimentally determining the wavefunctions in terms of the anisotropy of the 4f electrons remains challenging.

Figures 1G-1I show the VED distributions around the *A* = Pr, Nd, and Eu sites, respectively, obtained from the CDFS analysis of $A_2Ir_2O_7$ at 100 K. The local Cartesian coordinate system for the *A* site at (1/2, 1/2, 1/2) is defined as ***x*** || 2***a*** – ***b*** – ***c***, ***y*** || ***b*** – ***c***, and ***z*** || ***a*** + ***b*** + ***c***. Here, the *z*-axis is the local three-fold rotation axis and the *y*-axis is perpendicular to a mirror plane. Anisotropic VED is observed around the Pr site, as shown by yellow iso-density surfaces (Fig. 1G). Around the Nd site, the yellow iso-density surface appears relatively isotropic, while the orange surface, representing a higher density level, exhibits anisotropic distribution (Fig. 1H). In contrast to the two *A* ions, nearly isotropic VED is observed around the Eu site (Fig. 1I). In the case of $Eu^{3+}$ with $4f^6$ electrons, it is predicted that the total angular moment quantum number is *J* = 0 and the VED distribution is spherical, which is consistent with the experimental result.

To quantify the anisotropy of the 4f VED around the Pr and Nd sites, the density $\rho(\theta,\phi)$ at distances *r* = 0.33 and 0.24 Å from the nucleus, corresponding to the peak top of the radial profile of $\rho(\boldsymbol{r})$ along the *a*-axis (see Fig. S3), is shown by a color map on a sphere in Figs. 2A and 2B, respectively. The color scale shows $\left[\rho(\theta,\phi)-\overline{\rho(\theta,\phi)}\right]/\overline{\rho(\theta,\phi)}$. The anisotropy of VED is greater in $Pr^{3+}$ than in $Nd^{3+}$. The $4f^n$ ground state in the trigonal CEF can be described approximately as a linear combination of different $J_z$ eigenstates in the *LS* coupling scheme. The trigonal CEF $\hat{H}_{\mathrm{CEF}}$ can be expanded [36] as

$$\hat{H}_{\mathrm{CEF}} = B_{20}\hat{O}_{20} + B_{40}\hat{O}_{40} + B_{43}\hat{O}_{43} + B_{60}\hat{O}_{60} + B_{63}\hat{O}_{63} + B_{66}\hat{O}_{66}. \tag{1}$$

Here, $B_{lm}$ and $\hat{O}_{lm}$ are the CEF parameters and the CEF operators, respectively. In the case of $4f^2$ of $Pr^{3+}$, the *J* = 4 nonuplet is split into three doublets and three singlets, as shown in Fig. 1C. The doublets and singlets can be represented as

$$|\Gamma_\alpha \pm\rangle = C_1|J_z = \pm 4\rangle + C_2|J_z = \pm 1\rangle + C_3|J_z = \mp 2\rangle, (|C_1|^2 + |C_2|^2 + |C_3|^2 = 1) \tag{2}$$

and

$$|\Gamma_\beta \pm\rangle = C_4|J_z = \pm 3\rangle + C_5|J_z = 0\rangle + C_6|J_z = \mp 3\rangle, (|C_4|^2 + |C_5|^2 + |C_6|^2 = 1) \tag{3}$$

respectively. Here, double-sign corresponds. In the case of $4f^3$ of $Nd^{3+}$, the *J* = 9/2 decuplet is split into five Kramers doublets, as shown in Fig. 1D. Two doublets can be represented as

$$|\Gamma_\gamma \pm\rangle = C_1\left|J_z = \pm\frac{9}{2}\right\rangle + C_2\left|J_z = \pm\frac{3}{2}\right\rangle + C_3\left|J_z = \mp\frac{3}{2}\right\rangle, (|C_1|^2 + |C_2|^2 + |C_3|^2 = 1) \tag{4}$$

and the other three as

$$|\Gamma_\delta \pm\rangle = C_4\left|J_z = \pm\frac{7}{2}\right\rangle + C_5\left|J_z = \pm\frac{1}{2}\right\rangle + C_6\left|J_z = \mp\frac{5}{2}\right\rangle. (|C_4|^2 + |C_5|^2 + |C_6|^2 = 1) \tag{5}$$

where double-sign corresponds.

We optimize the coefficients $C_i$ (*i* = 1~3 or 4~6) to reproduce the anisotropy of $\rho(\theta,\phi)$ obtained by the CDFS analysis. The *R* value for the fitting of $\rho(\theta,\phi)$ is defined as

$$R = \frac{\sum_{\theta,\phi}|\rho(\theta,\phi) - s\rho_e(\theta,\phi;\Gamma)|}{\sum_{\theta,\phi}|\rho(\theta,\phi)|}. \tag{6}$$

Here, $\rho_e(\theta,\phi;\Gamma)$ is the square of the absolute value of the spherical harmonics term in the Γ state calculated by Eq. S1, as in Ref. [37], and *s* is the scale factor. The *R* values as a function of $\{C_i\}$ are shown as two-dimensional color maps: Figs. 2C and 2D for *A* = Pr, and Figs. 2G and 2H, for *A* = Nd. The optimized $\{C_i\}$ parameters are listed in Tables 1 and 2. The state with the lowest *R* value ($R_{\mathrm{min}}$) is shown by a surface color plot in Figs. 2E, 2F, 2I, and 2J. Around the Pr site, $\rho_e(\theta,\phi;\Gamma_\alpha+)$ (Fig. 2E) with the $R_{\mathrm{min}}$ value is in better agreement with the experimental VED $\rho(\theta,\phi)$ (Fig. 2A) than $\rho_e(\theta,\phi;\Gamma_\beta+)$ (Fig. 2F). The obtained $\{C_i\}$ parameters for the $\Gamma_\alpha+$ state are consistent with those of the lowest energy determined by the point charge model calculation considering SOC and trigonal CEF (Table 1). This finding is also consistent with previous inelastic neutron scattering measurements of $Pr_2Ir_2O_7$ [38]. Around the Nd site, $\rho_e(\theta,\phi;\Gamma_\gamma+)$ (Fig. 2I) is in better agreement with the experimental VED $\rho(\theta,\phi)$ (Fig. 2B) than $\rho_e(\theta,\phi;\Gamma_\delta+)$ (Fig. 2J), which is also consistent with the results of the point charge model calculation (Table 2). We conclude that the CDFS analysis is useful for visualizing spin-orbit entangled 4f electrons and for directly identifying the nature of their wavefunctions from the VED distributions.

**Discussion**

Finally, we comment on the $5d^5$ VED around the $Ir^{4+}$ site (Fig. 3A), which has the same $D_{3d}$ site symmetry as the *A* site. The 5d electrons around the Ir atom, which are influenced by relativistic SOC, affect the topological electronic state of this system [30-35]. Figures 3B-3D show the experimentally observed VED distributions, represented by yellow and orange iso-density surfaces at different density levels, around the Ir site of *A* = Pr, Nd, and Eu systems, respectively. Although the anisotropies of the VEDs are roughly similar to one another, the VEDs themselves are rather less smooth. Furthermore, negative VED appears near the Ir nucleus in all systems (Fig. S4B), which is unphysical. The rapid change and negative sign of VED around the Ir site may be caused by the absence of high-Q diffraction data and the narrow dynamic range of the obtained data (details are described in Supporting Information).

In summary, we have demonstrated the visualization of spin-orbit entangled 4f electrons around lanthanide ions in $A_2Ir_2O_7$ (*A* = Pr, Nd, and Eu) by the CDFS analysis using high-photon-energy X-ray diffraction experiments. The anisotropic VED distributions observed around the Pr and Nd sites contrast with the isotropic VED distribution around the Eu site. The $4f^2$ electrons around the Pr site exhibit greater anisotropy than the $4f^3$ electrons around the Nd site. These findings are well-explained by the *LS* coupling scheme incorporating SOC and trigonal CEF effects, where the anisotropy reflects the coefficients of the linear combination of $J_z$ eigenstates. The proposed 4f orbital observation method is applicable to various crystalline materials exhibiting exotic properties, such as diverse magnetic structures [6], multipolar orders [7], and heavy fermion states [10], and can directly reveal the 4f wavefunctions.

**Materials and Methods**

**Sample preparation**

Single crystals of $A_2Ir_2O_7$ (*A* = Pr, Nd, and Eu) were grown by the KF flux method. Firstly, polycrystalline samples of $A_2Ir_2O_7$ were prepared by solid-state reactions of rare-earth oxides ($Pr_6O_{11}$, $Nd_2O_3$, and $Eu_2O_3$) and iridate $IrO_2$. The materials with the prescribed molar ratios were ground, pressed into a pellet, and sintered at 1000 °C for several days. The sintered polycrystals were ground and mixed with KF in a ratio of 1:200. The mixtures were placed in a platinum crucible covered with a lid. The crucible was cooled down to 850 °C at a rate of 2 °C/hour following anneal at 1100 °C for 5 h. After cooling to room temperature, crystals were separated from the KF residual flux by rinsing with distilled water. Octahedral single crystals were obtained.

**X-ray diffraction experiments**

X-ray diffraction experiments were performed on BL02B1 at a synchrotron facility SPring-8, Japan [39]. An $N_2$-gas-blowing device was employed to cool the crystals to 100 K. A two-dimensional detector CdTe PILATUS, which had a dynamic range exceeding $10^6$, was used to record the diffraction pattern. The X-ray wavelength was calibrated to $\lambda$ = 0.30946 Å. The intensities of Bragg reflections of the interplane distance $d$ > 0.28 Å were collected by CrysAlisPro program [40] using a fine slice method, in which the data were obtained by dividing the reciprocal space region in increments of $\Delta\omega$ = 0.01°. Intensities of equivalent reflections were averaged and the structural parameters were refined using Jana2006 [41]. The X-ray diffraction experiments of $A_2Ir_2O_7$ (*A* = Pr, Nd, and Eu) detected no structural phase transitions down to 100 K, which was consistent with the previous X-ray diffraction experiments [25]. The structural analysis found no significant off-stoichiometry at each atomic site. Here, the structural refinement was performed using only high-angle reflections of $\sin\theta/\lambda$ > 0.6 $Å^{-1}$. Since the contribution of spatially spread valence electrons to X-ray diffraction is negligible in the high-angle region [17], the structural parameters including the atomic displacement parameters were obtained with high accuracy. The detailed structural parameters of $A_2Ir_2O_7$ are summarized in Tables S1-S6.

**Point-charge model calculations**

The CEF energies using the point charge model in the strong spin-orbit coupling limit were calculated by the software package PyCrystalField [42] based on the obtained crystal structures.

**CDFS analysis**

The CDFS method [15] was used to extract the VED distribution around each atomic site at 100 K. [He]-, [Xe]-, and ([Xe] $4f^{14}$)-type electron configurations were regarded as core electrons for O, lanthanide (Pr, Nd, and Eu), and Ir atoms, respectively. The effect of the thermal vibration was subtracted from the VED using the atomic displacement parameters determined by the high-angle analysis. The voxel size of the three-dimensional electron density distribution was set to be $\Delta V = (0.02\ \text{Å})^3$. It should be noted that the absolute value of the obtained VED does not directly reproduce the number of valence electrons around the atoms because the effect of double scattering, absorption, extinction, and detector saturation could not be completely excluded in the measurement of diffraction intensities. Crystal structure and ED distribution were visualized by using VESTA [43].

**DFT calculations**

To obtain the atomic form factors for the CDFS analysis, fully relativistic all-electron calculations based on the density functional theory (DFT) were performed for isolated O, Pr, Nd, Eu, and Ir atoms using Quantum ESPRESSO [44]. The Perdew-Burke-Ernzerhof functional [45] was employed to approximate the exchange-correlation functional. The resulting radial distribution functions of each orbital of Pr, Nd, Eu, and Ir atoms are summarized in Fig. S2.

**Acknowledgments**

We thank H. Kusunose for fruitful discussions. This work was supported by JSPS KAKENHI (Grant Nos. 22K14010, 23K03307, 23H04869, 24H01644, 24H01649, and 24K00582) and JST FOREST (Grant No. JPMJFR2362). Support was also provided by the RIKEN TRIP initiative (RIKEN Quantum, Advanced General Intelligence for Science Program, Many-body Electron Systems). The synchrotron radiation experiments were carried out at SPring-8 with the approval of the Japan Synchrotron Radiation Research Institute (JASRI) (Proposal No. 2022A1751, 2022B1582, 2023A1687, and 2023B1603).

## Figures and Tables

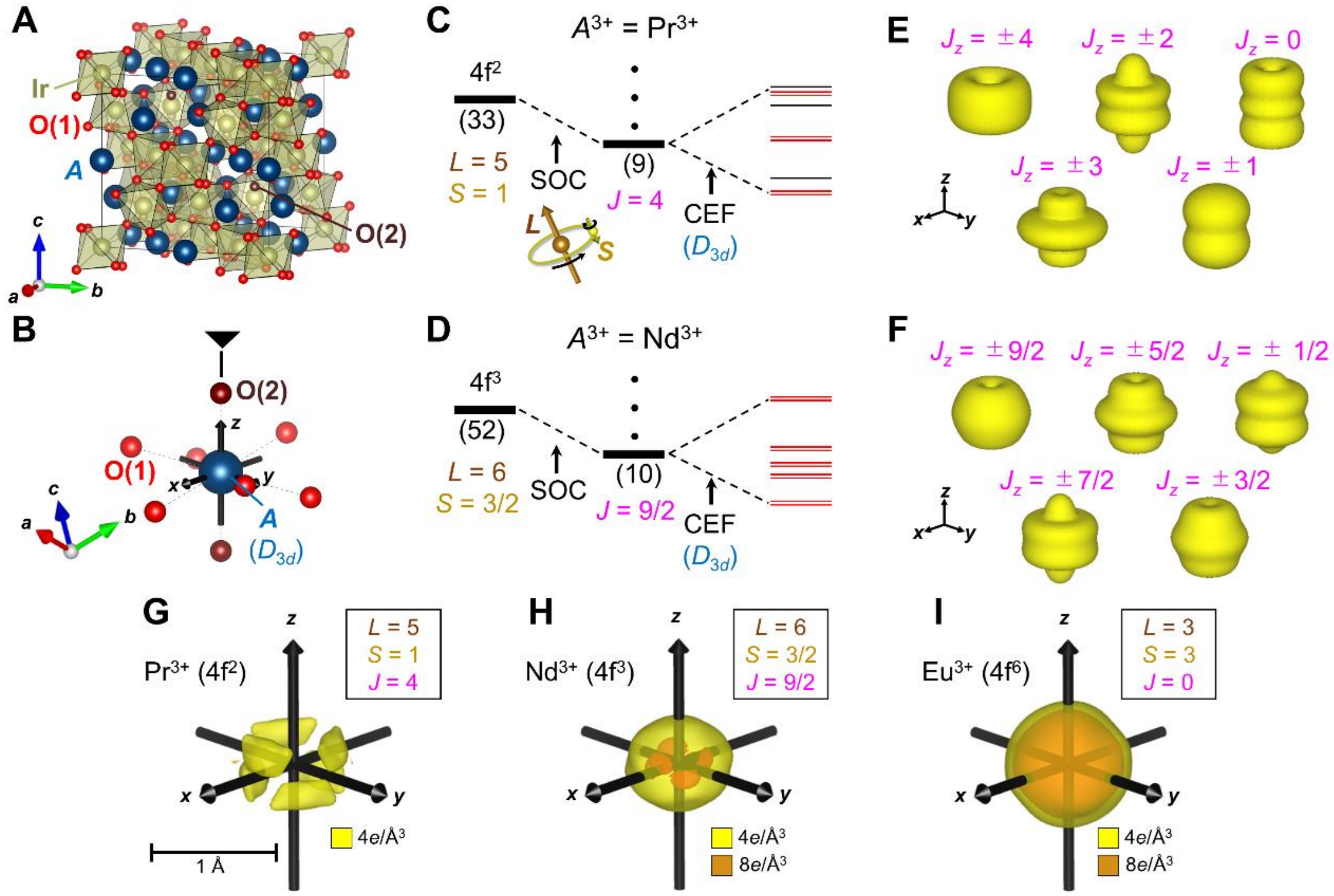

**Figure 1. Crystal structure of $A_2Ir_2O_7$ and spin-orbit entangled 4f state on the *A* site.** (A) Crystal structure of $A_2Ir_2O_7$. (B) *A* atom surrounded by eight O atoms: six O(1) and two O(2). A solid triangle indicates local three-fold rotation axis. (C,D) Schematic of $4f^2$ and $4f^3$ states considering the SOC and trigonal CEF under the *LS* coupling scheme, respectively. Single black and double red lines in the right-most diagram represent singlets and doublets, respectively. (E,F) $4f^2$ and $4f^3$ electron distributions of each $J_z$ eigenstate in the $^3H_4$ ($J = 4$, $L = 5$, $S = 1$) and $^4I_{9/2}$ ($J = 9/2$, $L = 6$, $S = 3/2$) multiplets, respectively. (G-I) Experimentally obtained VED distributions around the Pr, Nd, and Eu sites, respectively. Yellow and orange iso-density surfaces show electron-density levels of 4 and 8 $e$/Å$^3$, respectively.

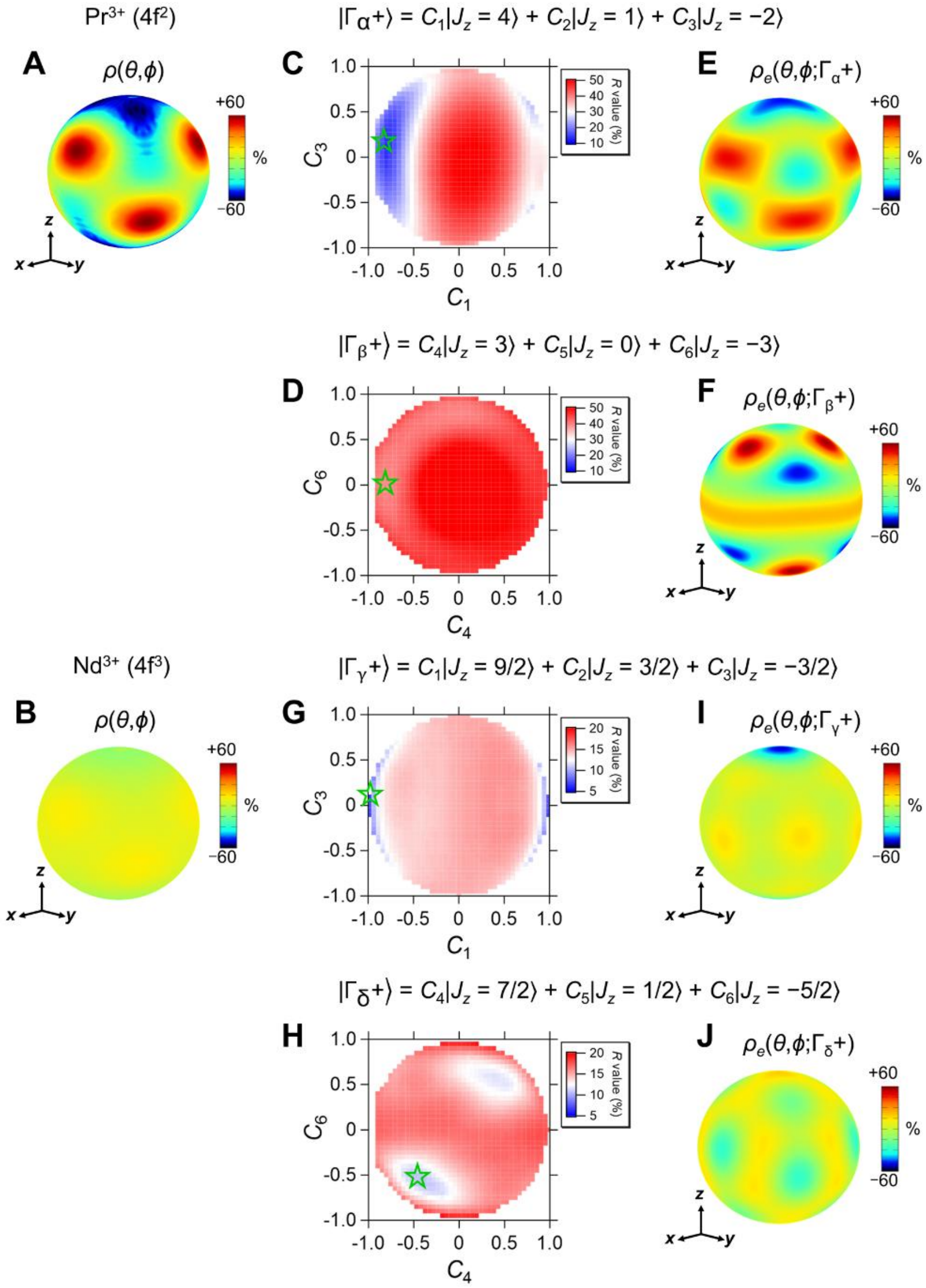

**Figure 2. VED distributions around the *A* site.** (A,B) Direction dependence of VED distributions $\rho(\theta,\phi)$ at a distance $r$ = 0.33 and 0.24 Å from the Pr and Nd sites, respectively, obtained by the CDFS analysis. The color scale indicates $\left[\rho(\theta,\phi)-\overline{\rho(\theta,\phi)}\right]/\overline{\rho(\theta,\phi)}\times 100$ [%]. (C,D) Color maps of $R$ values on two-dimensional $C_1$—$C_3$ ($C_2 \geq 0$) and $C_4$—$C_6$ ($C_5 \geq 0$) planes assuming the $|\Gamma_\alpha+\rangle$ and $|\Gamma_\beta+\rangle$ states of $4f^2$ electrons on a $Pr^{3+}$ ion in the trigonal field, respectively. (E,F) Simulated surface color plots of $\rho_e(\theta,\phi;\Gamma_\alpha+)$ and $\rho_e(\theta,\phi;\Gamma_\beta+)$, respectively. (G,H) Color maps of $R$ values on two-dimensional $C_1$—$C_3$ ($C_2 \geq 0$) and $C_4$—$C_6$ ($C_5 \geq 0$) planes assuming the $|\Gamma_\gamma+\rangle$ and $|\Gamma_\delta+\rangle$ states of $4f^3$ electrons on an $Nd^{3+}$ ion in the trigonal field, respectively. (I,J) Simulated surface color plots of $\rho_e(\theta,\phi;\Gamma_\gamma+)$ and $\rho_e(\theta,\phi;\Gamma_\delta+)$, respectively. The sets of coefficients $\{C_i\}$ in (E, F, I, and J) are shown by green stars in (C, D, G, and H), respectively. The color bar scale is plotted as $[\rho_e(\theta,\phi;\Gamma)-N_e]/N_e\times100$ [%]. Here, $N_e$ is the number of 4f electrons.

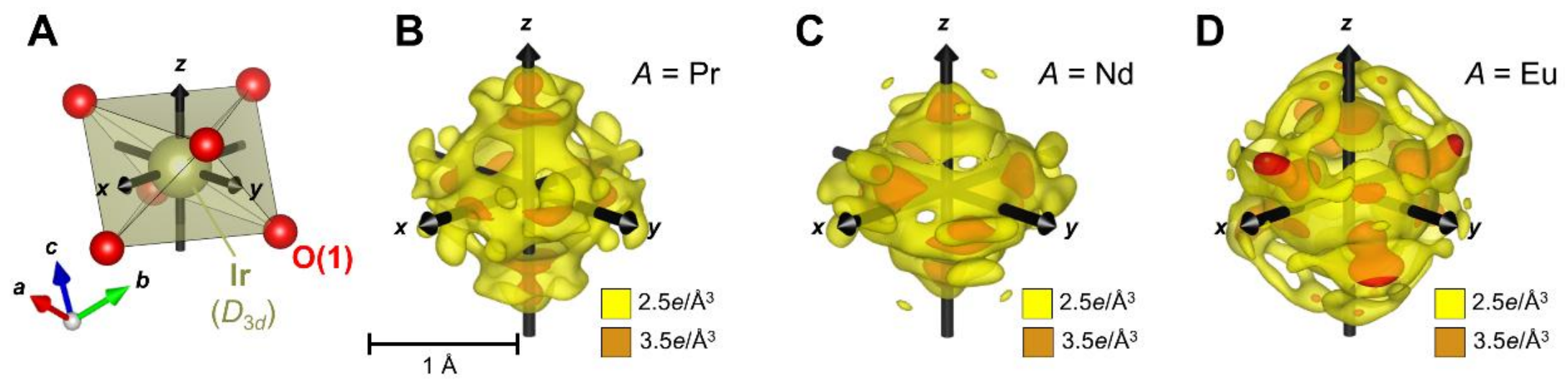

**Figure 3. 5d$^5$ VED around the Ir$^{4+}$ site.** (A) An Ir atom surrounded by six O(1) atoms, forming a $IrO_6$ octahedron. (B-D) Experimentally obtained VED distributions around the Ir site in *A* = Pr, Nd, and Eu systems, respectively. Yellow and orange iso-density surfaces show electron-density levels of 2.5 and 3.5 *e*/Å$^3$, respectively.

**Table 1.** Summary of the fitting analysis of $4f^2$ VED and the point-charge model calculations for the ground state ($E$ = 0) of a $Pr^{3+}$ ion under the trigonal CEF. $|n\rangle$ denotes the state with $J$ = 4 and $J_z$ = $n$.

| State | $\|4\rangle$ | $\|3\rangle$ | $\|2\rangle$ | $\|1\rangle$ | $\|0\rangle$ | $\|-1\rangle$ | $\|-2\rangle$ | $\|-3\rangle$ | $\|-4\rangle$ | $R_{min}$ [%] |
|---|---|---|---|---|---|---|---|---|---|---|
| $\Gamma_{\alpha}+$ | -0.79 | 0 | 0 | 0.60 | 0 | 0 | 0.15 | 0 | 0 | 13.13 |
| $\Gamma_{\beta}+$ | 0 | -0.81 | 0 | 0 | 0.59 | 0 | 0 | 0.00 | 0 | 39.36 |
| Cal. ($E$ = 0) | -0.91 | 0 | 0 | 0.41 | 0 | 0 | 0.08 | 0 | 0 | |

**Table 2.** Summary of the fitting analysis of $4f^3$ VED and the point-charge model calculations for the ground state ($E$ = 0) of an $Nd^{3+}$ ion under the trigonal CEF. $|n\rangle$ denotes the state with $J$ = 9/2 and $J_z$ = $n$.

| State | $\|9/2\rangle$ | $\|7/2\rangle$ | $\|5/2\rangle$ | $\|3/2\rangle$ | $\|1/2\rangle$ | $\|-1/2\rangle$ | $\|-3/2\rangle$ | $\|-5/2\rangle$ | $\|-7/2\rangle$ | $\|-9/2\rangle$ | $R_{min}$ [%] |
|---|---|---|---|---|---|---|---|---|---|---|---|
| $\Gamma_{\gamma}+$ | -0.99 | 0 | 0 | 0.05 | 0 | 0 | 0.13 | 0 | 0 | 0 | 6.59 |
| $\Gamma_{\delta}+$ | 0 | -0.47 | 0 | 0 | 0.70 | 0 | 0 | -0.54 | 0 | 0 | 10.38 |
| Cal. ($E$ = 0) | -0.96 | 0 | 0 | 0.07 | 0 | 0 | -0.27 | 0 | 0 | 0 | |